\documentclass[prl,showpacs,twocolumn,floats,10pt,aps,citeautoscript,longbibliography,superscriptaddress]{revtex4-2}
\usepackage{dcolumn}
\usepackage{amsmath}
\usepackage{mathrsfs}
\usepackage{amsfonts}
\usepackage{braket} 
\usepackage[dvipsnames]{xcolor}
\usepackage{enumitem}
\usepackage{graphicx}
\usepackage{float}
\usepackage{tikz}
\usepackage{algorithm}
\usepackage{algpseudocode}

\usepackage{amsmath} 
\usepackage{amssymb}
\usepackage{mathrsfs} %needed for uppercase \ell: \eLL = \mathscr{L} 
\usepackage{bbm}  % double 1 for unit matrix \bbm{1}
\allowdisplaybreaks % page break in equations
\usepackage{enumitem} %to use enumerate environment with options

\usepackage{scalefnt} % needed for \scalefont{}

\RequirePackage[hyperindex,colorlinks,bookmarksnumbered, plainpages=true,pdfstartview=FitH]{hyperref}
\hypersetup{linkcolor=blue,urlcolor=blue,citecolor=blue}
\usepackage{hyperref}

\definecolor{dimgreen}{rgb}{0.2,0.7,0.2}
\definecolor{grey}{rgb}{.6,.6,.6}

\newcommand{\ellminusone}{{\ell \hspace{-0.2mm}-\hspace{-0.2mm} 1}}
\newcommand{\ellplusone}{{\ell  \hspace{-0.2mm}+\hspace{-0.2mm} 1}}

\newcommand{\ellb}{{\bar \ell}}
\newcommand{\ellp}{{\ell'}}

\newcommand{\eLL}{{\mbox{\small$\mathscr{L}$}}}
\newcommand{\seLL}{{\scriptscriptstyle \! \mathscr{L}}}

\def\Mc{\mathcal{M}}
\def\Pc{\mathcal{P}}
\def\Oc{\mathcal{O}}
\def\Ponesite{\Pc^{1\mathrm{s}}}

\newcommand{\shrewd}{shrewd}
\newcommand{\Shrewd}{Shrewd}
\newcommand{\oneTDVP}{1{TDVP}}
\newcommand{\twoTDVP}{2{TDVP}}
\newcommand{\onesite}{\textrm{1s}}
\newcommand{\twosite}{\textrm{2s}}

\newcommand{\monesite}{\mathrm{1s}}
\newcommand{\mtwosite}{\mathrm{2s}}
\newcommand{\mbond}{\mathrm{b}}

\def\Ab{{\overline{A}}}

\def\Bb{{\overline{B}}}

\def\Db{\overline{D}}

\def\Dt{\widetilde{D}} 
\def\Dh{\widehat{D}} 
\def\Dp{D^\prime} 
\def\Df{D_\fin}

\def\tmax{t_{\mathrm{max}}} 
\def\tb{\bar t} 

\def\epsilonp{\epsilon^\prime}
\def\epsilont{\widetilde \epsilon}
\def\epsilonf{\epsilon}

\def\xif{\xi}

\def\Lambdab{{\overline{\Lambda}}} 
\def\Lambdat{{\widetilde{\Lambda}}} 
\def\K{{\scriptstyle {\rm K}}} % discarded
\def\D{{\scriptstyle {\rm D}}} % discarded
\def\P{{\scriptstyle {\rm P}}} % discarded

\newcommand{\Dmax}{D_{\mathrm{max}}}

\newcommand{\trunc}{{\mathrm{tr}}}

\newcommand{\deff}{d_{\mathrm{eff}}}

\newcommand{\maximum}{{\mathrm{max}}}
\newcommand{\phonon}{{\mathrm{ph}}}
\newcommand{\omegaphonon}{\omega_\phonon}

\newcommand{\PE}{\Delta_{P}}

\newcommand{\fin}{{\mathrm{f}}}
\newcommand{\tot}{{\mathrm{tot}}}
\newcommand{\expand}{{\mathrm{ex}}}

\def\mi{\mathrm{i}}
\def\deltat{{\delta}}

\newcommand{\pdag}{{\protect\vphantom{dagger}}}

\newcommand{\EllipseWhiteLambda}{\protect\raisebox{0\height}{\protect\includegraphics[width=0.0333\linewidth]{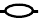}}}
\newcommand{\CircleWhiteC}{\protect\raisebox{-0.5mm}{\protect\includegraphics[width=0.037\linewidth]{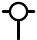}}}
\newcommand{\CircleGreenC}{\protect\raisebox{-0.5mm}{\protect\includegraphics[width=0.037\linewidth]{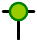}}}
\newcommand{\TriangleWhiteA}{\protect\raisebox{-0.5mm}{\protect\includegraphics[width=0.037\linewidth]{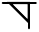}}}
\newcommand{\TriangleWhiteB}{\protect\raisebox{-0.5mm}{\protect\includegraphics[width=0.037\linewidth]{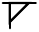}}}
\newcommand{\TriangleGreyA}{\protect\raisebox{-0.5mm}{\protect\includegraphics[width=0.037\linewidth]{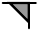}}}
\newcommand{\TriangleGreyB}{\protect\raisebox{-0.5mm}{\protect\includegraphics[width=0.037\linewidth]{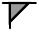}}}

\newcommand{\TriangleOrangeA}{\protect\raisebox{-0.5mm}{\protect\includegraphics[width=0.037\linewidth]{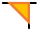}}}

\newcommand{\TriangleGreenA}{\protect\raisebox{-0.5mm}{\protect\includegraphics[width=0.037\linewidth]{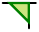}}}

\newcommand{\Atrunc}{{\widetilde A}{}^\trunc}

\newcommand{\Eq}[1]{Eq.~\eqref{#1}}

\makeatletter
\def\maketitle{
\@author@finish
\title@column\titleblock@produce
\suppressfloats[t]}
\makeatother

\begin{document}
\title{Time-dependent variational principle with controlled bond expansion for matrix product states}

\author{Jheng-Wei Li}
\affiliation{Arnold Sommerfeld Center for Theoretical Physics, Center for NanoScience,\looseness=-1\,  and Munich 
Center for \\ Quantum Science and Technology,\looseness=-2\, Ludwig-Maximilians-Universität München, 80333 Munich, Germany}
\author{Andreas Gleis}
\affiliation{Arnold Sommerfeld Center for Theoretical Physics, Center for NanoScience,\looseness=-1\,  and Munich 
Center for \\ Quantum Science and Technology,\looseness=-2\, Ludwig-Maximilians-Universität München, 80333 Munich, Germany}
\author{Jan von Delft}
\affiliation{Arnold Sommerfeld Center for Theoretical Physics, Center for NanoScience,\looseness=-1\,  and Munich 
Center for \\ Quantum Science and Technology,\looseness=-2\, Ludwig-Maximilians-Universität München, 80333 Munich, Germany}

\begin{abstract}
\begin{center}
(Dated: \today)
\end{center}
We present a controlled bond expansion (CBE) approach to simulate quantum dynamics based on the time-dependent variational principle (TDVP) for matrix product states.
Our method alleviates the numerical difficulties of the standard, fixed-rank one-site TDVP integrator by increasing bond dimensions on the fly to reduce the projection error. This is achieved in an economical, local fashion, requiring only minor modifications of  standard one-site TDVP implementations.
We illustrate the performance of CBE--TDVP with several numerical examples on finite quantum lattices.
\\
\\
\noindent
DOI:
\end{abstract}

\maketitle

\textit{Introduction.---}
The time-dependent variational principle (TDVP) \cite{dirac_1930,McLachlan1964,Meyer1990,RMP_1994} is a standard tool for time-evolving the Schr\"{o}dinger equation on a constrained manifold parametrizing the wave function.
Tensor networks (TN) offer efficient parametrizations based on low-rank approximations \cite{Verstraete2008,Cirac2009,Eisert2010,Schollwoeck2011,Stoudenmire2012,Bridgeman2017,Orus2019,Silvi2019}.
Their combination, TN--TDVP, holds much potential for studying the dynamics of quantum lattice models \cite{Koch2007,Koch2010,Haegeman2011,Koffel2012,Hauke2013,Lubich2013,Lubich2013a,Haegeman2013,Lubich2015a,Haegeman2016,Kieri2016,ZaunerStauber2018,Vanderstraeten2019,Bauernfeind2020,Rams2020,Secular2020,Bauernfeind2020,Kloss2020,Ceruti2021,Kohn2021,Damme2021}, quantum field theories \cite{Milsted2013,Gillman2017}, and quantum chemistry problems  \cite{Chin2016,Chin2018,Kurashige2018,Kloss2019,Xie2019,Xu2022}.

Here, we focus on matrix product states (MPSs), an elementary class of TN states. 
Their time evolution, pioneered in Refs.~\cite{Vidal2004,Daley2004,White2004}, can be treated using a variety of methods, reviewed in Refs.~\cite{Schollwoeck2011,Paeckel2019}. Among these, MPS--TDVP \cite{Haegeman2011,Lubich2013,Lubich2013a,Haegeman2013,Lubich2015a,Haegeman2016}, 
which uses Lie-Trotter decom\-po\-si\-tion to integrate a train of tensors sequentially, arguably gives the best results regarding both physical accuracy and performance \cite{Paeckel2019}:  it (i) is applicable for long-ranged Hamiltonians, 
and its one-site (\onesite) version (\oneTDVP) ensures (ii) unitary time evolution, (iii) energy conservation \cite{Hairer2006,Haegeman2011} and (iv) numerical stability \cite{Lubich2013,Lubich2015a,Kieri2016}.

A drawback of \oneTDVP, emphasized in Refs.~\onlinecite{Kloss2018,Goto2019,Chanda2020}, is use of a \textit{fixed}-rank integration scheme. 
This offers no way of dynamically adjusting the MPS rank (or bond dimension), as needed to track the entanglement growth typically incurred during MPS time evolution. 
For this, a rank-adaptive two-site (\twosite) TDVP (\twoTDVP) algorithm can be used \cite{Haegeman2016}, but it has much higher computational costs and in practice does not ensure properties (ii-iii).

To remedy this drawback, we introduce a rank-adap\-tive integrator for \oneTDVP\ that is more efficient than previous ones  \cite{Dektor2020,Yang2020,Ceruti2022,Dunnett2021}.
It ensures properties (i-iv) at the same numerical costs as \oneTDVP, with marginal overhead. 
Our key idea is to control the TDVP projection error \cite{Haegeman2016,Hubig2018,Dektor2020} by adjusting  MPS ranks on the fly via the controlled bond expansion (CBE) scheme of Ref.~\cite{Gleis2022}.
CBE finds and adds subspaces missed by \onesite\ schemes but containing significant weight from $H\Psi$. 
When used for DMRG ground state searches, CBE yields \twosite\ accuracy with faster convergence per sweep, at \onesite\ costs \cite{Gleis2022}. 
CBE--TDVP likewise comes at essentially \onesite\ costs.

\textit{MPS basics.---}
Let us recall some MPS basics, adopting the notation of Refs.~\onlinecite{Gleis2022a,Gleis2022}. 
For an $\eLL$-site 
system an open boundary MPS wave function $\Psi$ having dimensions $d$ for physical sites and $D$ for virtual bonds can 
always be written in site-canonical form,
\begin{align}
  \Psi & =  \, 
  \raisebox{-3.0mm}{\includegraphics[width=0.75\linewidth]{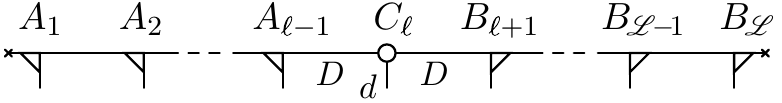}} \, .
  \label{eq:canonical}
\end{align}
The tensors $C_\ell\!$ (\CircleWhiteC), $A_\ell\!$  (\TriangleWhiteA) and $B_{\ell}\!$ (\TriangleWhiteB) are variational pa\-rameters. $A_\ell$ and $B_\ell$ are  left and right-sided isometries, respectively, projecting $Dd$-dimensional \textit{parent} 
($\P$) spaces to $D$-dimensional \textit{kept} ($\K$) images spaces; they obey
\vspace{-2mm}
\begin{flalign}
\label{eq:IsometricConditions}
& \raisebox{-6.5mm}{
 \includegraphics[width=0.883\linewidth]{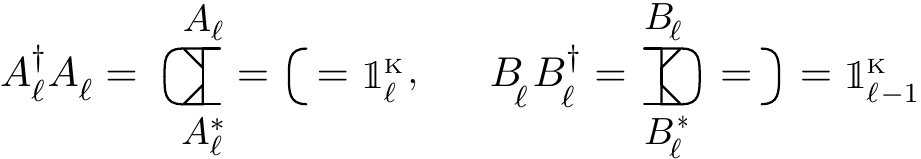}} .  \hspace{-1cm} &
 \end{flalign}
The gauge relations $C_\ell  \! = \! A_\ell \Lambda_\ell \!=\! \Lambda_{\ellminusone} B_\ell$ ensure that \Eq{eq:canonical} remains unchanged when moving the orthogonality center $C_\ell$ from one site to another.

The Hamiltonian can likewise be expressed as a matrix product operator (MPO) with virtual bond dimension $w$,
\vspace{-4mm}
\begin{align}
H = \raisebox{-3.0mm}{\includegraphics[width=0.75\linewidth]{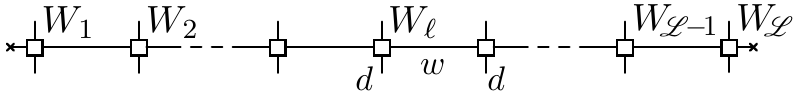}}  \, . &
\end{align}
Its projection to the effective local state spaces associated with site $\ell$  or bond $\ell$ yields effective one-site or zero-site Hamiltonians, respectively, computable recursively via 
\vspace{-1mm}
\begin{subequations}
  \label{subeq:H1H0}
\begin{align}
  H_{\ell}^\monesite  & = \raisebox{-5mm}{\includegraphics[width=0.72\linewidth]{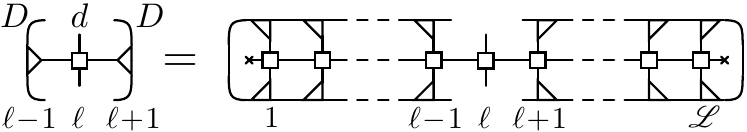}} \,\, , 
  \label{eq:H_i}
  \\ 
  H_{\ell}^\mbond & = 
  \raisebox{-5mm}{\includegraphics[width=0.566\linewidth]{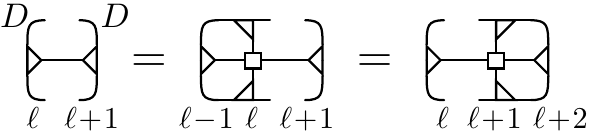}} \!\! . 
  \label{eq:K_i}
\end{align}
\end{subequations}
These act on \onesite\ or bond representations of the wave function, $\psi^\monesite _\ell \!=\! C_\ell (\CircleWhiteC)$ or $\psi^\mbond_\ell \!=\! \Lambda_\ell (\EllipseWhiteLambda)$, respectively. 

Let $\Ab_\ell\!$ (\TriangleGreyA) and $\Bb_\ell\!$ (\TriangleGreyB) 
be isometries that are or\-tho\-go\-nal complements
of $A_\ell$ and $B_\ell$,  with \textit{discarded} ($\D)$ image spaces of dimension $\Db \!=\! D(d \!-\! 1)$, obeying orthonor\-ma\-li\-ty \pagebreak 
\noindent 
and completeness relations complementing Eq.~\eqref{eq:IsometricConditions} \cite{Gleis2022}: 
\vspace{-1mm}
\begin{subequations}
\label{subeq:AdditionalOrthonormalityRelations}
\begin{flalign}
\label{eq:AdditionalOrthonormalityRelations}
 &
 \raisebox{-4mm}{
 \includegraphics[width=0.89\linewidth]{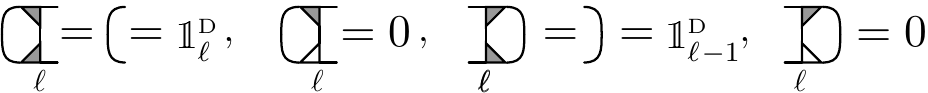}} ,
 \hspace{-1cm} &
\\
\label{eq:Completeness_Main}
 & \hspace{-1mm} \raisebox{-3.2mm}{
 \includegraphics[width=0.886\linewidth]{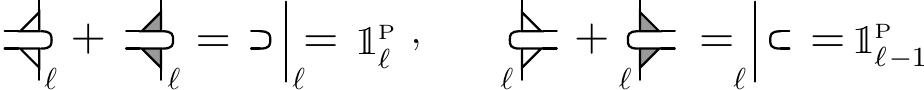}} .
 \hspace{-2cm} &
\end{flalign}
\end{subequations}

\vspace{-1mm}
\textit{Tangent space projector.---}
Next, we recapitulate the TDVP strategy.
It aims to solve the Schr\"odinger equation, 
$\mi \dot \Psi \!=\! H \Psi$, constrained to the manifold $\Mc$ of all MPSs of the form \eqref{eq:canonical}, with \textit{fixed} bond dimensions. 
Since $H\Psi$ typically has larger bond dimensions than $\Psi$ and hence does not lie in $\Mc$, the TDVP aims to minimize $\| \mi \dot \Psi \!-\! H \Psi \|$ within $\Mc$. This leads to 
\vspace{-1mm}
\begin{align}
\mi \dot \Psi(t) = \Ponesite (t) H \Psi (t),
\label{eq:TDVP}
\end{align}

\vspace{-1mm}
\noindent 
where $\Ponesite (t)$ is the projector onto the tangent space of $\Mc$ at $\Psi(t)$, i.e.\ the space of all \onesite\ variations of $\Psi (t)$:
\vspace{-1mm}
\begin{flalign}
\label{eq:tangentspaceprojector}
& \Ponesite = \raisebox{-3.5mm}{\includegraphics[width=0.623\linewidth]{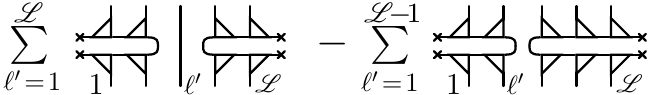}}
\\
& \; = \raisebox{-3.5mm}{\includegraphics[width=0.903\linewidth]{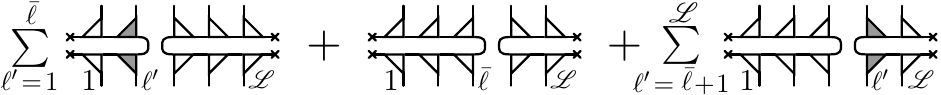}}\,  .
\hspace{-1cm} &
\nonumber
\end{flalign}
The form in the first line was derived by Lubich, Oseledts, and Vandereycken~\cite{Lubich2015a} (Theorem 3.1), and transcribed into MPS notation in Ref.~\cite{Haegeman2016}. 
For further explanations of its form, see Refs.~\cite{Gleis2022a,supplement}.
The second line, valid for any $\ellb \!=\! 1, \dots, \eLL\!-\! 1$, follows via \Eq{eq:Completeness_Main}; \Eq{eq:AdditionalOrthonormalityRelations}
implies that all its terms conveniently are mutually orthogonal, and that the projector property $(\Ponesite)^2 = \Ponesite$ holds \cite{Gleis2022a}.

\iffalse
The structure \eqref{eq:tangentspaceprojector} of $\Ponesite$ can motivated by the following short-cut argument (equivalent to invoking gauge invariance \cite{Haegeman2016}).
If $\Psi$ has the MPS form \eqref{eq:canonical}, then $\dot \Psi$ is a sum of terms each containing one derivative. 
Bond $\ell$, represented by $C_\ell B_\ellplusone \!=\! A_\ell \Lambda_\ell B_\ellplusone$, yields 
$\dot A_\ell \Lambda_\ell B_\ellplusone \!+\!  A_\ell \dot \Lambda_\ell B_\ellplusone \!+\! A_\ell \Lambda_\ell \dot B_\ellplusone$. 
Decomposing $\dot A_\ell$ as $A_\ell \Lambda'_\ell + \Ab_\ell \Lambdab{}'_\ell$ and $\dot B_\ellplusone$ as $\Lambda^{\prime \prime}_\ell B_\ellplusone + \Lambdab{}^{\prime\prime}_\ell \Bb_\ellplusone$, 
we obtain the structure
\vspace{-1mm}
\begin{align}
\raisebox{-2.5mm}{\includegraphics[width=0.883\linewidth]{Eq/LocalVariationOfPsi.pdf}} ,
\end{align}
with $\Lambdat_\ell =\Lambda'_\ell + \dot \Lambda_\ell + \Lambda^{\prime \prime}_\ell$.
$\Ponesite H$ must yield terms of match\-ing structure, 
i.e.\ $\Ponesite$ must contain the combination
\vspace{-2mm}
\begin{align}
\label{eq:partialtangentspaceprojector}
\Bigl[ \raisebox{-3.7mm}{\includegraphics[width=0.38\linewidth]{Eq/LocalVariationProjectorSimplified.pdf}} \Bigr] \, , 
\end{align}

\vspace{-2mm}
\noindent
for any $\ell \! \in \!  [1, \eLL\!-\!1]$. Indeed it does,
as seen from the second line of \eqref{eq:tangentspaceprojector}
with the choices $\ellb \!=\! \ell$, and  
$\ellp\!=\!\ell$ in the first term, 
$\ellp\!=\ell\!+\!1$ in the third.
\fi

\textit{One-site TDVP.---}
The \oneTDVP\ algorithm \cite{Lubich2015a,Haegeman2016} represents \Eq{eq:TDVP}  by $2 \eLL\!-\!1$ coupled equations, $\mi \dot C_\ell \!=\!  H^\monesite _\ell C_\ell$ and $\mi \dot \Lambda_\ell \!=\! - H^\mbond_\ell \Lambda_\ell$, stemming, respectively, from the $\eLL$ single-site and $\eLL\!-\!1$ bond projectors 
of $\Ponesite$ (\Eq{eq:tangentspaceprojector}, first line). 
Evoking a Lie-Trotter decomposition, they are then decoupled and for each time step solved sequentially, for $C_\ell$ or $\Lambda_\ell$ (with all other tensors fixed).
For a time step from $t$ to $t' \! = \! t  + \deltat$
one repeatedly performs four substeps, e.g.\ sweeping right to left:
(1) Integrate $\mi \dot{C}_{\ellplusone} \!=\! H_{\ellplusone}^\monesite C_{\ellplusone}$ from $t$ to $t'$; 
(2) QR factorize $C_{\ellplusone} (t') \!=\! \Lambda_{\ell} (t') B_{\ellplusone}(t')$;
(3) integrate $\mi \dot{\Lambda}_{\ell} \!=\! - H_{\ell}^\mbond \Lambda_{\ell}$ from $t'$ to $t$; and 
(4) update $A_{\ell} (t) C_{\ellplusone} (t) \to C_\ell (t) B_\ellplusone (t')$, with $C_\ell(t) = A_{\ell} (t) \Lambda_{\ell} (t)$. 
 
\oneTDVP\ has two leading errors.
One is the Lie-Trotter decomposition error.  
It can be  reduced by higher-order integration schemes \cite{McLachlan1995,Hairer2006}; we use a third-order integrator with error $\Oc(\deltat^3)$ \cite{HigherOrderIntegrators}.
The second error is the projection error from projecting the Schr\"{o}dinger equation into the tangent space of $\Mc$ at $\Psi(t)$, quantified 
by $\PE \!=\! \|(\mathbbm{1} \!-\! \Ponesite) H \Psi (t) \|^2$. 
It can be reduced brute force by increasing the bond dimension, as happens when using \twoTDVP\ \cite{Haegeman2016,Paeckel2019,Goto2019}, or through global subspace expansion \cite{Yang2020}. 
Here, we propose a  local approach, similar in spirit to that of Ref.~\cite{Dunnett2021}, but more efficient, with \onesite\ costs, and without stochastic ingredients, in contrast to \cite{Xu2022}.

\textit{Controlled bond expansion.---}
Our key idea is to use CBE to reduce the \twosite\ contribution in $\PE$, given by 
$\PE^{2\perp} \!=\! 
\bigl\| \Pc^{2\perp} H \Psi 
\bigr \|^2$, where $\Pc^{2\perp} =\Pc^\mtwosite (1 \! - \! \Ponesite)$.
Here, $\Pc^\mtwosite$ is the projector onto \twosite\ variations
of $\Psi$, and $\Pc^{2\perp}$ its component orthogonal to the tangent space projector (see also \cite{Gleis2022a}):
\vspace{-3mm}
\begin{subequations}
\begin{flalign}
\label{eq:orthogonal-two-site-projector}
 \Pc^\mtwosite & = 
 \raisebox{-3.15mm}{\includegraphics[width=0.616\linewidth]{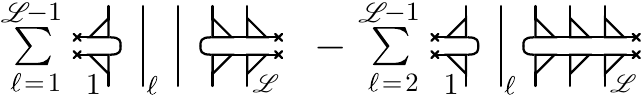}} \, , 
 \\  \Pc^{2\perp}
 & = \raisebox{-3.15mm}{\includegraphics[width=0.266\linewidth]{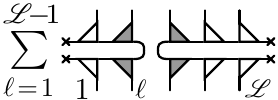}} \, , \quad 
\label{eq:TwoSiteVariance}
\PE^{2\perp}  = \!
\raisebox{-5.75mm}{\includegraphics[width=0.3\linewidth]{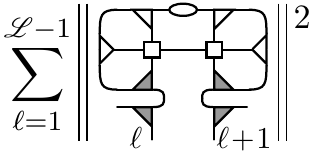}} \! . \hspace{-1cm} &
\end{flalign}
\end{subequations}

\vspace{-2mm}
\noindent
Now note that $\PE^{2\perp}$ 
is equal to
$\Delta_{E}^{2\perp} =  \|\Pc^{2\perp}(H \! - \! E)\Psi \|^2$, 
the \twosite\ con\-tri\-bution to the energy variance  
\cite{Hubig2018,Gleis2022,Gleis2022a}. 
In Ref.~\cite{Gleis2022}, discussing ground state searches via CBE--DMRG, we showed how to minimize $\Delta_{E}^{2\perp}$  at \onesite\ costs:
each bond $\ell$ can be expanded in such a manner that the added subspace carries significant weight from $\Pc^{2\perp}H\Psi$.
This expansion removes that subspace from the image of $\Pc^{2\perp}$, thus reducing $\Delta_{E}^{2\perp}$ significantly. 
Consider, e.g., a right-to-left sweep and let $\Atrunc_\ell$ (\TriangleOrangeA) be a truncation of $\Ab_\ell$ (\TriangleGreyA) having an image spanning such a subspace, of dimension $\Dt$, say. To expand  bond $\ell$ from $D $ to $D + \Dt$, we replace $A_\ell (\TriangleWhiteA)$ by $A_\ell^\expand (\TriangleGreenA)$, $C_\ellplusone (\CircleWhiteC)$ by $C_\ellplusone^{\expand} (\CircleGreenC)$
and $H^\monesite_\ellplusone$ by $H^{\monesite,\expand}_\ellplusone$,
with expanded tensors defined as 
 \vspace{-2mm}
\begin{flalign}
\label{eq:A1expand}
& \!  
\raisebox{-4mm}{
 \includegraphics[width=0.9\linewidth]{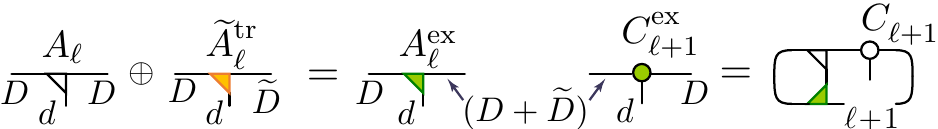}} \!\! ,
\hspace{-1cm} & 
\\ 
\label{eq:H1expand}
& \qquad H^{(1,\expand)}_\ellplusone  =
\raisebox{-4.7mm}{
\includegraphics[width=0.516\linewidth]{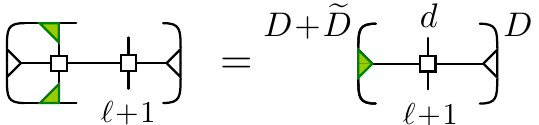}} \, . & 
\end{flalign}
Note that $\Psi$ remains unchanged, $A^\expand_\ell C_\ellplusone^{\expand} = A_\ell C_\ellplusone $.

Similarly, the projection error $\PE^{2\perp}$ can be minimized through a suitable choice of the truncated complement $\Atrunc_\ell (\TriangleOrangeA)$ \cite{Gleis2022}.
We find $\Atrunc_\ell$ using the so-called
\textit{\shrewd\ selection}
strategy of Ref.~\cite{Gleis2022} (Figs.~1 and 2 there); it avoids computation of \TriangleGreyA,\TriangleGreyB\, and has \onesite\ costs regarding 
CPU and memory, thus becoming increasingly advantageous for large $D$ and $d$.
\Shrewd\ selection involves two truncations ($D\!\to \! \Dp$ and $\Dh\! \to \! \Dt$ in Ref.~\cite{Gleis2022}). 
Here, we choose these to respect singular value thresholds of $\epsilonp \!=\! 10^{-4}$ and $\epsilont \!=\! 10^{-6}$, respectively; empirically, these yield good results in the benchmark studies presented below.

\textit{CBE--TDVP.---}
It is straightforward to incorporate CBE into the \oneTDVP\ algorithm:
simply expand each bond $\ell$ from $D\!\to \!D \!+\! \Dt$ before time-evolving it.  
Concretely, when sweeping right-to-left, we add step~(0): expand $A_\ell,C_\ellplusone,H^\monesite_\ellplusone \to A^\expand_\ell,C^\expand_\ellplusone,H^{\monesite,\expand}_\ellplusone$ following Eq.~\eqref{eq:A1expand} (and by implication also $\Lambda_\ell, H^\mbond_\ell \to \Lambda^\expand_\ell, H^{\mbond,\expand}_\ell$).
The other steps remain as before, except that in (2) we replace the QR factorization by an SVD. 
This allows us to reduce (trim) the bond dimension from $D+ \Dt$ to 
a final value $\Df$, as needed in two situations \cite{Dektor2020,Vanhecke2021,Ceruti2022}:
First, while standard \oneTDVP\ requires keeping and even padding small singular values in order to retain a fixed bond dimension \cite{Koch2007,Lubich2013}, that is not necessary here.
Instead, for bond trimming, we discard small singular values below an empirically determined threshold $\epsilonf=10^{-12}$. 
This keeps the MPS rank as low as possible, without impacting the accuracy \cite{Dektor2020}.
Second, once $D +\Dt$ exceeds $D_\maximum$, we trim it back down to $D_\maximum$ aiming to limit computational costs.
The trimming error is characterized by its discarded weight, $\xif(t)$, which we either control or monitor. The TDVP properties of (ii) unitary evolution and (iii) energy conservation \cite{Ceruti2022} hold to within order  $\xi(t)$.

\begin{figure}
  \includegraphics[width=1.0\linewidth,trim = 0.0in 0.0in 0.0in 0.0in,clip=true]{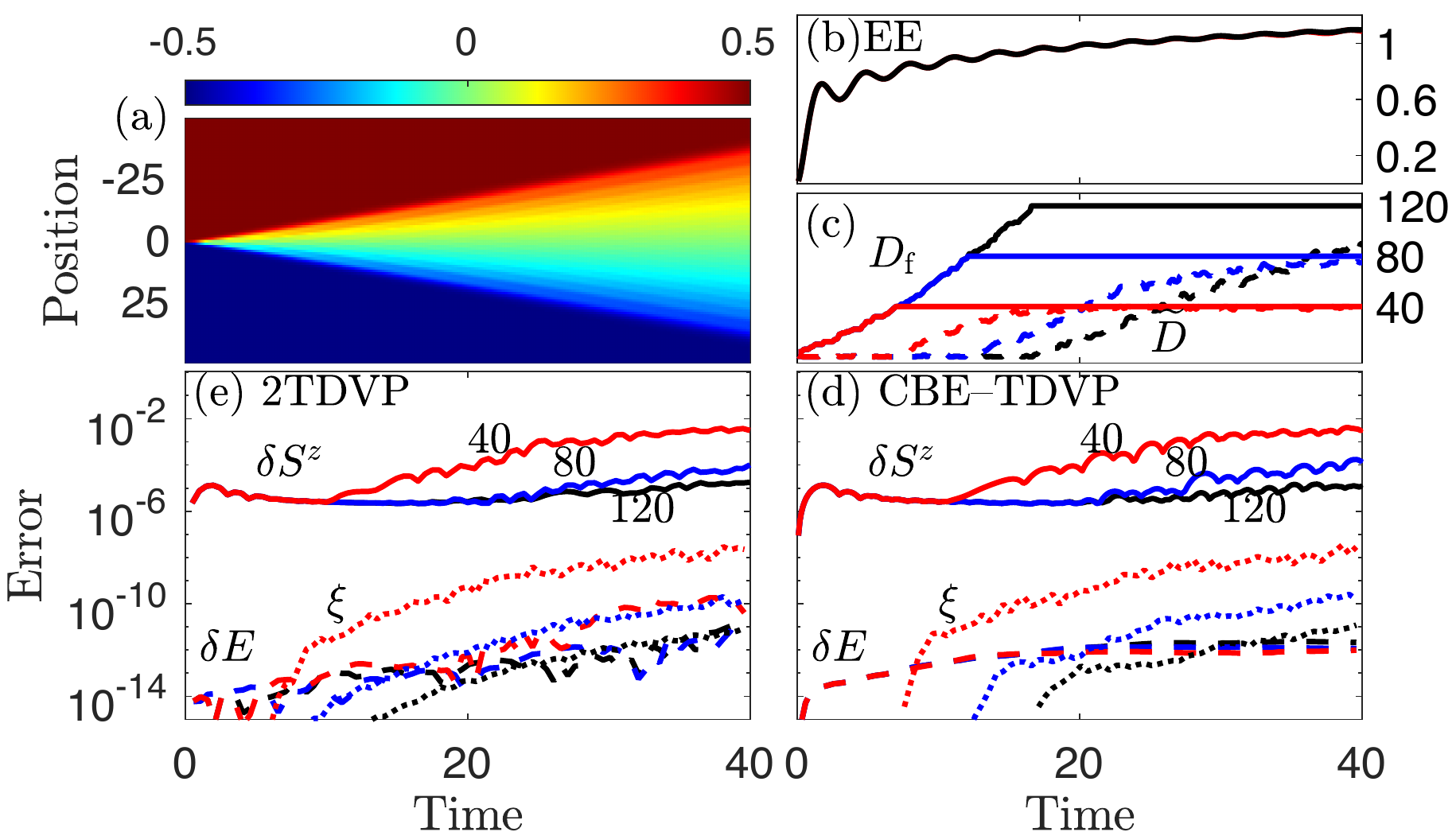} \vspace{-5mm}
  \caption{100-site XX spin chain: Time evolution of a domain wall, computed with time step $\deltat=0.05$ and U$(1)$ spin symmetry.
  (a) Local magnetization profile $S^{z}_{\ell}(t)$.
  (b) Entanglement entropy EE$(t)$ between the left and the right half of the chain.
  (c) Bond dimension $\Df(t)$ and its pre-trimming expansion $\Dt(t)$ per time step, for $\Dmax\!=\!120$.
  (d,e) Error analysis: magnetization $\delta S^{z}(t)$ (solid line),i.e., the maximum deviation (over $\ell$) of $S_\ell^z(t)$ from the exact result,
, energy $\delta E(t)$ (dashed line), and discarded weight $\xif(t)$ (dotted line) for $\Dmax\!=\!40$ (red), $80$ (blue) and $120$ (black), computed with (d) CBE--TDVP or (e) 2TDVP. Remarkably, the errors are comparable in size, although CBE--TDVP has much smaller computational costs.
  } \vspace{-5mm}
  \label{fig:XY}
\end{figure}

\textit{Results.---} 
We now benchmark CBE--TDVP for three spin models, then illustrate its perfor\-mance 
for large $d$ using the Peierls--Hubbard model with $d=36$.
Our bench\-mark comparisons track the time evolution of the entanglement entropy EE($t$) between the left and right halves of a chain, the bond dimensions $\Df(t)$ and $\Dt(t)$, the discarded weight $\xi(t)$, the deviations from exact results of spins expectation values, $\delta S(t)$, and the energy change, $\delta E(t)$, which should vanish for unitary time evolution.

\textit{XX model: domain wall motion.---} 
We consider a spin chain with Hamiltonian $H_\mathrm{XX}=\sum_{\ell}(S^{x}_{\ell}S^{x}_{\ell+1} + S^{y}_{\ell}S^{y}_{\ell+1})$. 
We compute the time evolution of the local magnetization profile $S^{z}_{\ell}(t) = \braket{\Psi (t)|\widehat{S}_\ell^z| \Psi (t)}$, initialized with a sharp domain wall, $\ket{\Psi(0)} \!=\! \ket{\uparrow\uparrow\!\ldots\!\uparrow\downarrow\downarrow\!\ldots\!\downarrow}$. 
For comparison, the analytical solution for $\eLL \to \infty$ reads \cite{Antal1999}
$ S^{z}_{\ell}(t) = -1/2 \sum_{n = 1 - \ell}^{\ell - 1} J_n(t)^2$,
for $\ell\geq 1$ (right half) and $S^{z}_{\ell} = -S^{z}_{1-\ell}$ otherwise, where $J_n(t)$ is the Bessel function of the first kind.
The domain wall spreads with time [Fig.~\ref{fig:XY}(a)], entailing a steady growth of the entanglement entropy (EE) between the left and right halves of the spin chain [Fig.~\ref{fig:XY}(b)].
$D(t)$ and $\Dt(t)$ [Fig.~\ref{fig:XY}(c)] start from 1 and 0.
Initially, $\Dt$ remains remarkably small ($\lesssim 10$), while $\Df$ increases in steps of $\Dt$ until reaching  $\Dmax$. Thereafter $\widetilde{D}$ increases noticeably, but remains below $\Dmax$ for all times shown here. 
This reflects CBE frugality---bonds are expanded only as much as needed.

\begin{figure}
  \includegraphics[width=1.0\linewidth,trim = 0.0in 0.0in 0.0in 0.0in,clip=true]{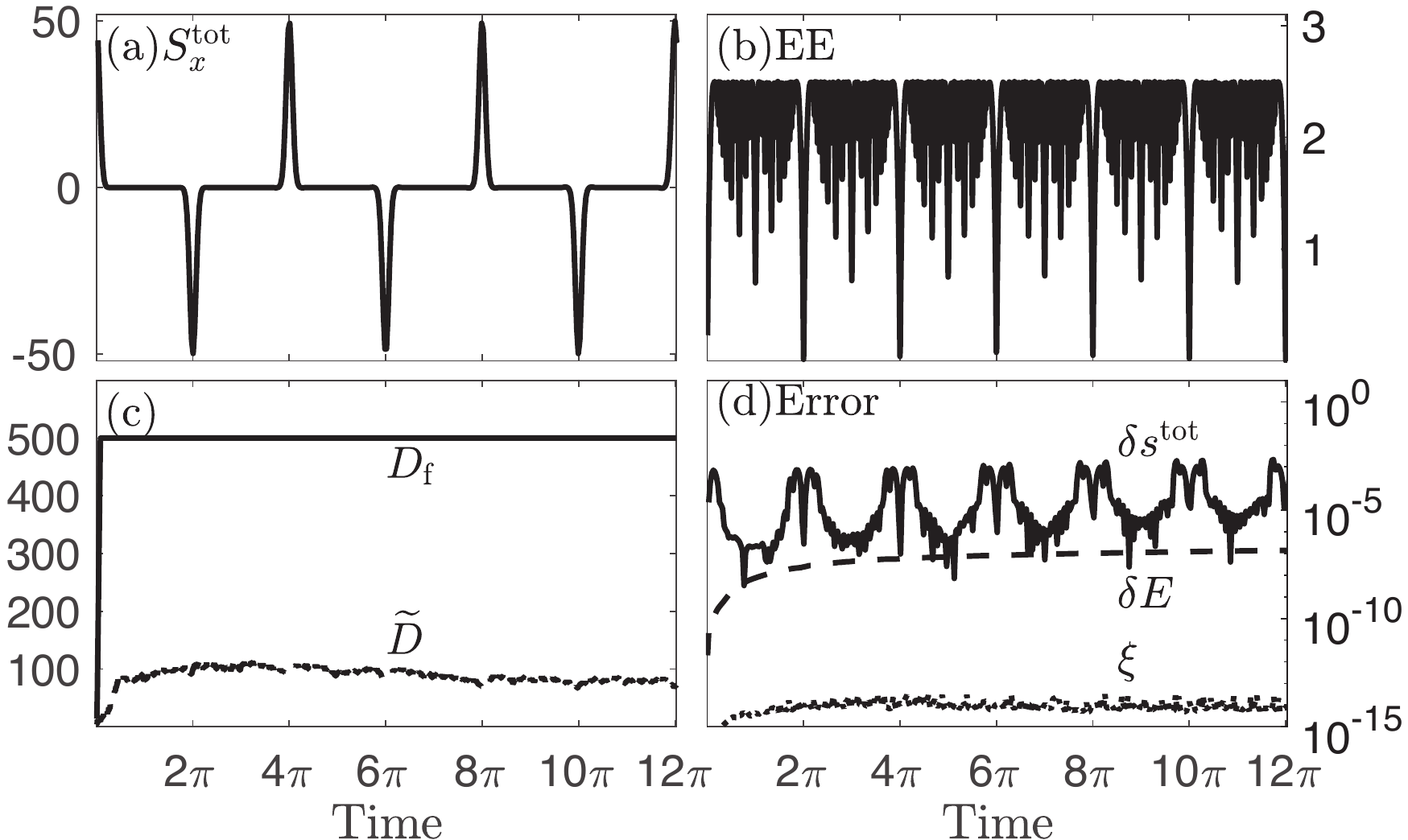}
  \vspace{-4mm}
  \caption{100-site one-axis twisting model: Time evolution 
  of an initially $x$-polarized spin state, computed using $\deltat=0.01$ and $\mathbb{Z}_2$ spin symmetry.
  (a) Total spin $S^{\rm{tot}}_x(t)$, (b) entanglement
  entropy, and (c) bond dimensions.
  (d) Error analysis: error in total spin density $\delta s^{\rm{tot}}_x(t)$ (solid line), energy $\delta E(t)$ (dashed line), and discarded weight $\xif(t)$ (dotted line), for $\Dmax=500$.
  } \vspace{-5mm}
  \label{fig:OAT}
\end{figure}

Figure~\ref{fig:XY}(d) illustrates the effects of
changing $\Dmax$, following the error analysis of Ref.~\onlinecite{Gobert2005}.
 The leading error is quantified by $\delta S^z(t)$
(solid line), the maximum deviation (over $\ell$) of 
$S_\ell^z(t)$ from the exact result. Comparing the data for $\Dmax=40, 80, 120$, we observe a finite bond dimension effect:
The error $\delta S^{z}$ increases appreciably once the discarded weight $\xif$ (dotted line) becomes larger than $10^{-11}$.
By contrast, the energy change (dashed line) stays small irrespective of the choice of $\Dmax$. (For more discussion of error accumulation, 
see Ref.~\cite{supplement}.) 
Figure~\ref{fig:XY}(e) shows a corresponding error analysis
for 2TDVP, computed using $D\!=\!\Dmax$;
its errors are comparable to those of 
CBE--TDVP, though the latter is much cheaper.

\textit{One-axis twisting (OAT) model: quantum revivals.---}
The OAT model has a very simple Hamiltonian, $H_\mathrm{OAT}=(\sum_{\ell}S^{z}_\ell)^2/2$, but its long-range interactions are a challenge for tensor network methods using real-space parametrizations. 
We study the evolution of
$S^\tot_x (t) = 
\bra{\Psi(t)}\sum_{\ell}\widehat{S}^{x}_{\ell}
\ket{\Psi(t)}$, for an initial  $\ket{\Psi(0)}$ having all spins $x$-polarized (an MPS with $D=1$).
The exact result, $S^{\rm{tot}}_x(t) = (L/2) {\rm{cos}}^{L-1}(t/2)$, exhibits periodic collapses and revivals \cite{Dooley2014}.
Yang and White \cite{Yang2020} have studied the short-time dynamics using TDVP with global subspace expansion, reaching times $t\leq0.5$.
CBE--TDVP is numerically stable for much longer times [Fig.~\ref{fig:OAT}(a)]; it readily reached $t = 12 \pi$, completing three cycles. 
(More would have been possible
with \textit{linear} increase in computation time.) 
This stability is remarkable, since the rapid initial growth of the entanglement entropy, the finite time-step size, and the limited bond dimension [Fig.~\ref{fig:OAT}(b,c)] cause some inaccuracies, which remain visible throughout [Fig.~\ref{fig:OAT}(d)].
However, such numerical noise evidently does not accumulate over time and does not spoil the long-time dynamics: 
CBE--TDVP retains the treasured properties (i-iv) of \oneTDVP, up to the truncation tolerance governed by $\xif$.

\begin{figure}
  \includegraphics[width=1.0\linewidth,trim = 0.0in 0.0in 0.0in 0.0in,clip=true]{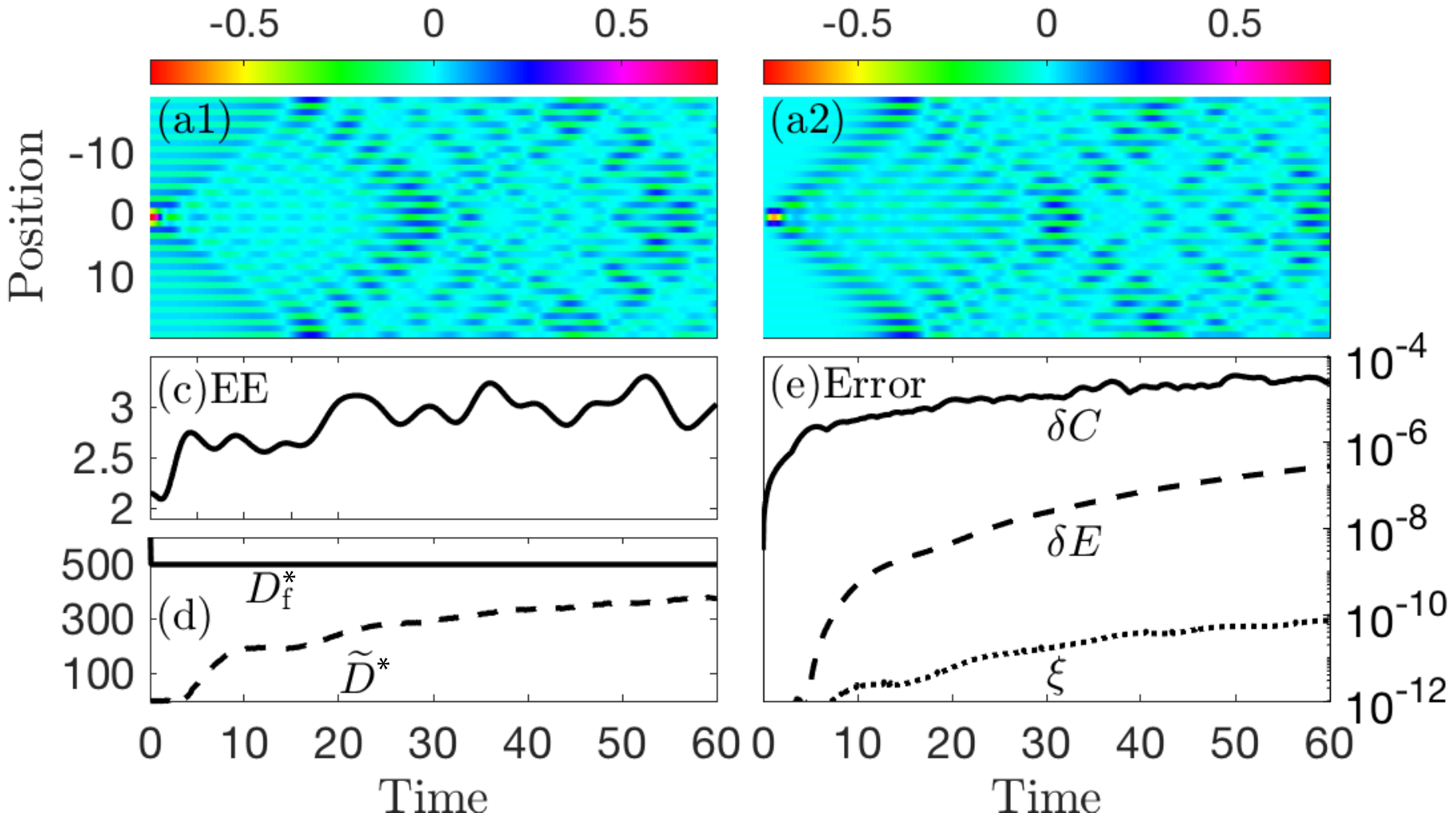} \vspace{-3mm}
  \includegraphics[width=1.0\linewidth]{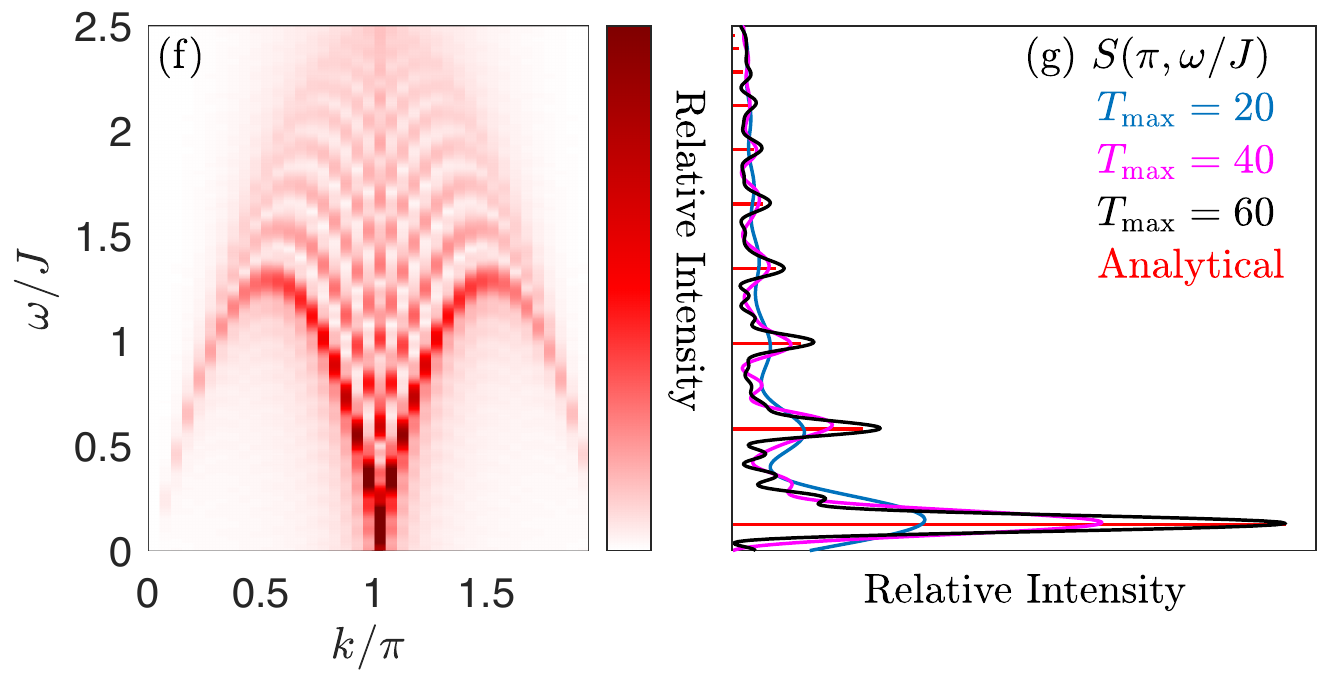}
  \caption{40-site SU(2) Haldane-Shastry model: (a-d) Time evolution of a spin excitation, computed with $\deltat=0.05$ and SU$(2)$ spin symmetry.   (a1,a2) Real and imaginary parts of $C(x,t)$,
  (b) entanglement entropy, and (c) bond dimensions. 
  (d) Error analysis: $\delta C(t)$, the maximum of $\delta C(x,t)$ over $x$ (solid line), energy $\delta E(t)$ (dashed line) and discarded weight $\xif(t)$ (dotted line), for $\Dmax=500$.
  (f) Normalized spectral function $S(k,\omega)/S(\pi,0)$, obtained using $t_\maximum = 60$.
  (g) $S(\pi,\omega)/S(\pi,0)$, obtained using $t_\maximum =20,40$, $60$; red lines indicate exact peak heights. 
  } \vspace{-7mm}
  \label{fig:HS}
\end{figure}

\textit{$\rm{SU}(2)$ Haldane-Shastry model: spectral function.---}
Our final benchmark example is the SU$(2)$ Haldane-Shastry model
on a ring of length $\eLL$, with Hamiltonian 
\vspace{-4mm}
\begin{align}
H_\textrm{HS} = J\sum_{\ell<\ell'\leq \seLL} \frac{\pi^2{\bf{S}}_{\ell}\cdot{\bf{S}}_{\ell'} }{\eLL^2{\rm{sin}}^2\frac{\pi}{\seLL}(\ell-\ell')}.
\label{Eq:HS2}
\end{align}

\vspace{-2mm}
\noindent
Its ground state correlator, $C(x,t) \!=\! 
\braket{\Psi_0|\widehat{\bf{S}}_x(t) \widehat{\bf{S}}_0(0)| 
\Psi_0}$,
is related by discrete Fourier transform 
to its spectral function, $S(k,\omega)$, given by ($0<\ell'<\ell\leq \eLL/2$) 
\cite{Yamamoto2000,Yamamoto2000b} 
\vspace{-2mm}
\begin{flalign}
& S \! \left(
2(\ell+\ell')\tfrac{\pi}{\seLL},
\tfrac{\pi^2}{\seLL^2}((\ell+\ell')\eLL-2(\ell^2+{\ell'}^2)+\ell-\ell')
\right) \hspace{-1cm} & 
\label{Eq:HS}
& \\
& \quad =
\frac{2\ell-2\ell'-1}{(2\ell-1)(\eLL-2\ell'-1)}
\prod_{\overline{\ell}=\ell'+1}^{\ell-1}
\frac{2\overline{\ell}(\eLL-2\overline{\ell})}{(2\overline{\ell}-1)(\eLL-2\overline{\ell}-1)}. \hspace{-1cm} &
\nonumber \end{flalign}
\vspace{-2mm}

Figures~\ref{fig:HS}(a,b) show the real and the imaginary parts of $C(x,t)$, computed using CBE--TDVP.
For early times ($t\leq20$), the  local excitation introduced at $\ell=0$, $t=0$ spreads ballistically, 
as reported previously \cite{Haldane1993,Zaletel2015,Secular2020}.
Once the counter-propagating wavefronts meet on the ring, an interference pattern emerges. 
Our numerical results remain accurate throughout, as shown by the error analysis in Fig.~\ref{fig:HS}(e). Figure~\ref{fig:HS}(f) shows the corresponding spectral function $S(k,\omega)$, obtained 
by discrete Fourier transform of $C(x,t)$ using a maximum simulation time of $t_\maximum$. Figure~\ref{fig:HS}(g) shows 
a cut along $k\!=\!\pi$: peaks
can be well resolved by increasing $t_\maximum$, 
with relative heights in excellent agreement with the exact 
\Eq{Eq:HS}.

\begin{figure}
  \includegraphics[width=1.0\linewidth,trim = 0.0in 0.0in 0.0in 0.0in,clip=true]{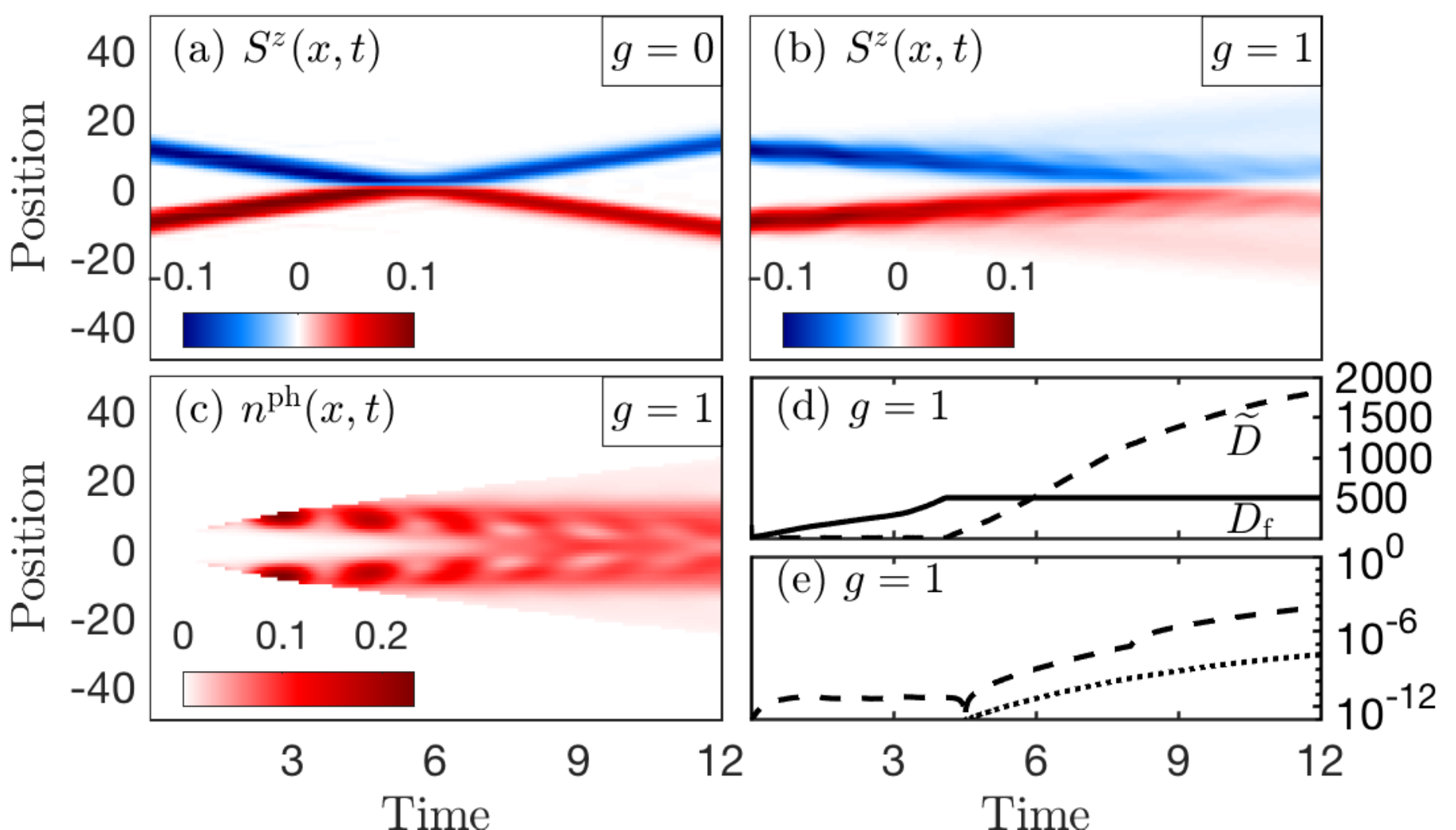} \vspace{-5mm}
  \caption{Peierls--Hubbard model: Real-space scattering of two electron wave packets, for $U\!=\!10$ and $\omegaphonon\!=\!3$,
  computed with $\deltat=0.05$, 
  $n_\maximum^\phonon \!=\!=8$ and U(1) spin symmetry.
   (a,b) Spin magnetic moment $S^z(x,t)$ for $g\!=\!0$ and $g\!=\!1$.
   (c) Phonon density $n^\phonon(x,t)$, (d) bond dimensions,  and (e) error analysis: energy $\delta E(t)$ (dashed line) and 
  discarded weight $\xif(t)$ (dotted line), all computed 
  for $g\!=\!1$, with $\Dmax=500$.
    } \vspace{-5mm}
  \label{fig:PFH}
\end{figure}

\textit{Peierls--Hubbard model:\! scattering dynamics.---}
Finally, we consider the scattering dynamics of interacting electrons coupled to phonons. This interaction leads to non-trivial
low-energy physics involving polarons \cite{Trugman2000,Clay2005,Fehske2007,Hague2007,Hardikar2007,Tezuka2007,Fehske2008,Marchand2010,Hohenadler2013,Hohenadler2016,Sous2018,Reinhard2019,Nocera2021};
the numerical study of polaron dynamics is currently attracting
increasing attention \cite{Fehske2007,Goleifmmode2012,Werner2013,Chen2015,Fetherolf2020,Pandey2021}.
Here, we consider the 1-dimensional Peierls--Hubbard model, 
\vspace{-2mm}
\begin{flalign} 
\label{eq:HubbardPeierls}
H_\textrm{PH} 
&  = 
\sum_{\ell}
U  n_{\ell\uparrow}n_{\ell\downarrow}
+ \sum_\ell \omegaphonon b^{\dagger}_{\ell}b^\pdag_{\ell} 
\hspace{-1cm} &
\\ \nonumber
& \; \; + 
\sum_{\ell\sigma}(c^{\dagger}_{\ell\sigma}c^\pdag_{\ellplusone \sigma} \!+\!
\mathrm{h.c.}) \bigl( -t + b^{\dagger}_{\ell} \!+\! b^\pdag_{\ell} 
\!-\! b^{\dagger}_{\ellplusone} 
\!-\! b_{\ellplusone} \bigr) \, . 
\end{flalign}

\vspace{-2mm}
\noindent 
Spinful electrons with onsite interaction strength $U$ 
and hopping amplitude $t\!=\!1$, 
and local phonons with frequency $\omegaphonon$,
are coupled with strength $g$ through a Peierls term modulating
the electron hopping. 

We consider two localized wave packets with opposite spins, average momenta $k=\pm\pi/2$ and width $W$ \cite{AlHassanieh2008,Moreno2013}, initialized as 
$\ket{\Psi_\pm} = \sum_\ell Ae^{-(\frac{x_{\ell} \mp x_{0}}{W})^2} e^{\mp \mi k x_{\ell}} c_{\ell \pm}^\dagger \ket{0}$, where $\ket{0}$ describes an empty lattice. 
Without electron-phonon coupling [$g=0$, Fig.~\ref{fig:PFH}(a)], there is little dispersion effect through the time of flight, and the strong interaction causes an elastic collision.
By contrast, for a sizable coupling 
in the nonperturbative regime \cite{Sous2018,Nocera2021}
[$g \!=\!1$, Figs.~\ref{fig:PFH}(b-e)], phonons are excited
by the electron motion [Fig.~\ref{fig:PFH}(c)].
After the two electrons have collided,
they show a tendency to remain close  to each other
(though a finite distance apart, since $U$ is large)
[Fig.~\ref{fig:PFH}(b)]; they thus
seem to form a bi-polaron, stabilized by 
a significant phonon density in the central region
[Fig.~\ref{fig:PFH}(c)]. 

We limited the phonon occupancy to $n^\phonon_\maximum\!=\!8$ per site. Then, $d= 4(n^\phonon_\maximum\!+\!1)\!=\!36$, and $\Db\!= \! 35 \Df$ is so large that \twoTDVP\ would be utterly unfeasible. 
By contrast, CBE--TDVP requires a comparatively small bond expansion of only $\widetilde{D}(t)\leq{4\Dmax}$ for the times shown; after that, the discarded weight $\xif(t)$ becomes substantial [Figs.~\ref{fig:PFH}(d,e)].

\textit{Conclusions and outlook.---}
Among the schemes for MPS time evolution, 1TDVP has various advantages (see introduction), but its projection error is uncontrolled.
2TDVP remedies this, albeit at \twosite\ costs, $\mathcal{O}(d^2 w D^3)$, and is able to simulate dynamics reliably \cite{Paeckel2019}.
CBE--TDVP at \onesite\ costs, $\mathcal{O}(d w D^3)$ achieves the same accuracy as 2TDVP.
Moreover, CBE--TDVP comes with significantly slower growth of bond dimensions $D$ in time, which speeds up the calculations further (see Ref.~\cite{supplement}).

Our benchmark tests of CBE--TDVP, on three exactly solvable spin models (two with long-range interactions), demonstrate its reliability. 
Our results on the Peierls--Hubbard model suggest that bi-polarons form during electron scattering---an effect not previously explored numerically.
This illustrates the potential of CBE--TDVP for tracking complex dynamics in computationally very challenging models.

For applications involving the time evolution of MPSs defined on ``doubled'' local state spaces, with effective local bond dimensions $\deff = d^2$,  the cost reduction of CBE--TDVP vs.\ 2TDVP, $\Oc(d^2 w D^3)$ vs.\ $\Oc(d^4 w D^3 )$, will be particularly dramatic. 
Examples are finite temperature properties, treated by purification of the density matrix \cite{Verstraete2004b} or dissipation-assisted operator evolution \cite{Rakovszky2022};
and the dynamics of open quantum systems \cite{Weimer2021}, described by
Liouville evolution of the density matrix \cite{Lindblad1976,Verstraete2004,Zwolak2004} or by  
an influence matrix approach \cite{Lerose2021}.

\bigskip
We thank Andreas Weichselbaum and Frank Pollmann for stimulating discussions, and Seung-Sup Lee, Juan Espinoza, Matan Lotem, Jeongmin Shim and
Andreas Weichselbaum for helpful comments on our manuscript.
 Our computations employed the QSpace tensor library \cite{Weichselbaum2012,Weichselbaum2020}. This research was funded in part by the Deutsche Forschungsgemeinschaft under Germany's Excellence Strategy EXC-2111 (Project No.\ 390814868), and is part of the  Munich Quantum Valley, supported by the Bavarian state government through the Hightech Agenda Bayern Plus. 

\vspace{-5mm}
\bibliography{CBE-TDVP}
\clearpage

\title{Supplemental material: Time-dependent variational principle with controlled bond expansion for matrix product states}

\date{\today}

\maketitle
\setcounter{secnumdepth}{2} % enable section numbering, which is disabled in prl
\renewcommand{\thefigure}{S-\arabic{figure}}% change figure numbering style for appendix
\setcounter{figure}{0}
\setcounter{section}{0}
\setcounter{equation}{0}
\renewcommand{\thesection}{S-\arabic{section}}% change section numbering style for appendix
\renewcommand{\theequation}{S\arabic{equation}}% change equation numbering style for appendix

\section{Single site (fixed rank) tangent space projector}
The structure \eqref{eq:tangentspaceprojector} of the tangent space projector $\Ponesite$ can be motivated by the following short-cut argument (equivalent to invoking gauge invariance \cite{Lubich2015a, Haegeman2016}).
If $\Psi$ is represented as an MPS, then its tangent vectors $\delta\Psi$ under the fixed-rank approximation can be expressed as a sum of MPSs each containing one derivative of a local tensor. 
This representation is not unique, but its gauge redundancy can be easily removed.
To do so, let us first consider the variation of MPS in Eq.~\eqref{eq:canonical} on a single bond $\ell$, i.e., $A_\ell C_\ellplusone \!=\! A_\ell \Lambda_\ell B_\ellplusone$, while the other tensors remain fixed (and hence are not depicted below). 
Its first order variation then gives us $\delta A_\ell \Lambda_\ell B_\ellplusone +  A_\ell \delta \Lambda_\ell B_\ellplusone + A_\ell \Lambda_\ell \delta B_\ellplusone$. 
By further rewriting $\delta A_\ell \Lambda_\ell$ as $A_\ell \Lambda'_\ell + \Ab_\ell \Lambdab{}'_\ell$ and $\Lambda_\ell \delta B_\ellplusone$ as $\Lambda^{\prime \prime}_\ell B_\ellplusone + \Lambdab{}^{\prime\prime}_\ell \Bb_\ellplusone$, 
we obtain the following unique decomposition,
\vspace{-1mm}
\begin{flalign}
& \raisebox{-2.5mm}{\includegraphics[width=0.883\linewidth]{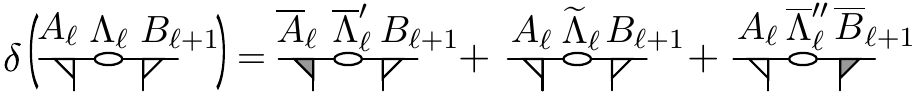}} , \hspace{-1cm} &
\end{flalign}
with $\Lambdat_\ell =\Lambda'_\ell + \delta \Lambda_\ell + \Lambda^{\prime \prime}_\ell$. 
The three terms on the right are mutually orthogonal to each other. Each of them belongs to the image space of one of the following three orthogonal projectors:
\vspace{-2mm}
\begin{align}
\label{eq:partialtangentspaceprojector}
\raisebox{-3.2mm}{\includegraphics[width=0.433\linewidth]{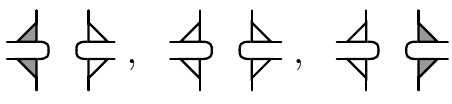}} \, ; 
\end{align}
their sum is a tangent space projector for $A_\ell \Lambda_\ell B_\ellplusone$.
Repeating the same argument for all the bonds, while avoiding double counting, i.e., including every term only once, we readily obtain $\Ponesite$ given by the second line of Eq.~\eqref{eq:tangentspaceprojector}.

Therefore, given an MPS of the form \eqref{eq:canonical}, $\Ponesite$ is indeed the orthogonal projector onto its tangent space under the fixed-rank approximation.
For real-time evolution, applying the Hamiltonian to $\ket{\Psi}$ leads the state out of its tangent space.
In the 1TDVP scheme, $H\ket{\Psi}$ is approximated by  $\Ponesite H\ket{\Psi}$, its orthogonal projection onto the tangent space,  leading to Eq.~\eqref{eq:TDVP}.

\section{Analysis of CBE-TDVP error propagation}

The TDVP time evolution of an MPS under the fixed-rank approximation is unitary, with energy conservation if the Hamiltonian is time-independent.
Expanding the tangent space does not spoil these desirable properties, provided
that no truncations are performed. However, then the bond dimension would keep growing with time, which is not practical for studies of long-time dynamics.

With our CBE approach, we instead restrict the bond dimension growth by  
bond trimming using $\epsilon=10^{-12}$, and also stopping the increase of $\Df$ once it has reached a specified maximal value $\Dmax$. 
Due to these truncations, the desirable TDVP properties are no longer satisfied exactly. However, {\it{for each time step}} they do hold within the truncation error, as shown by Ceruti, Kusch, and Lubich \cite{Ceruti2022}. Thus, the time evolution per time step is almost unitary. Nevertheless, errors can accumulate with time, hence it is unclear \textit{a priori} to what extent the desirable TDVP properties survive over long times.

\begin{figure}[t]
\vspace{-6mm}

\includegraphics[width=\linewidth]{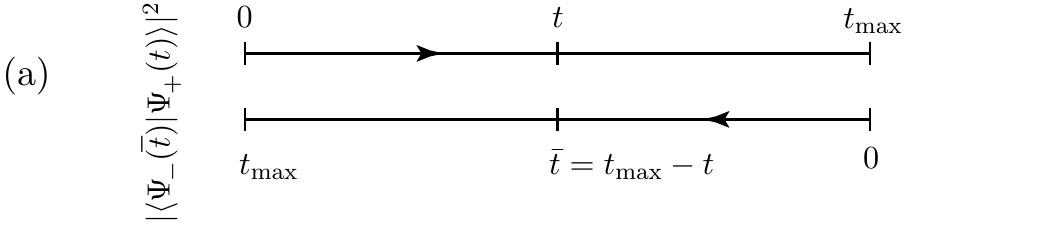} \qquad
\includegraphics[width=1.0\linewidth,trim = 0.0in 0.0in 0.0in 0.0in,clip=true]{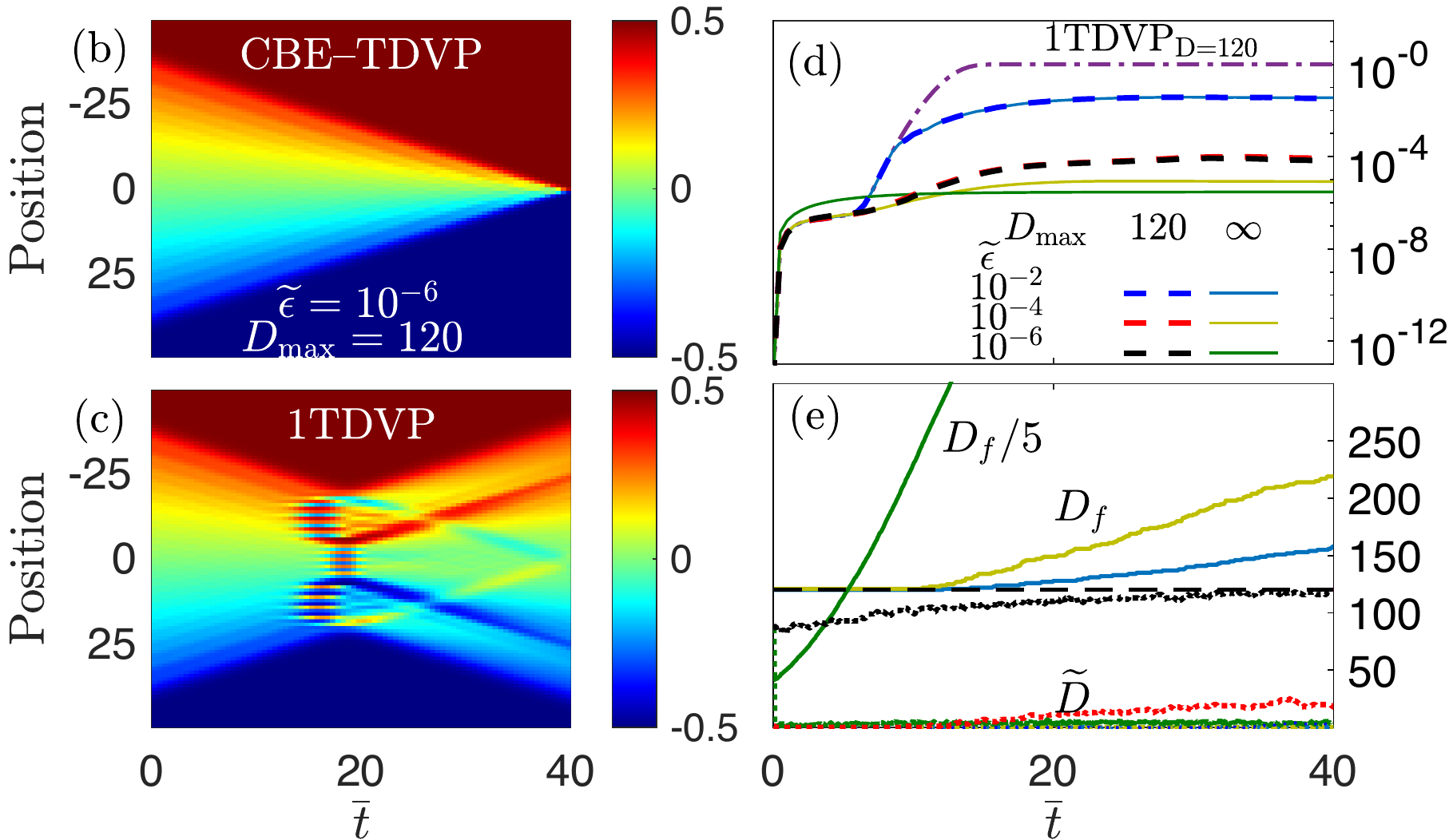}
\vspace{-6mm}

\caption{(a) Forward-backward time evolution for the computation of $F(t)$. 
  (b,c) Back-evolution of the domain wall, described by $\ket{\Psi_-(\tb)}$, computed using (b) CBE--TDVP and (c) \oneTDVP.
  (d) Time evolution of $\delta F(\tb) = 1-F(\tb)$, computed via \oneTDVP\ with $D=120$ (dash-dotted line), and via CBE--TDVP using three values of $\widetilde{\epsilon}$, and either with $D_{\rm{max}} \!=\! 120$ (dashed lines) or 
  $D_{\rm{max}} \!=\! \infty$ (solid lines).
  (e) Time evolution of the corresponding  bond dimensions $\Df (\tb)$ (solid lines) and $\Dt (\tb)$ (dots). 
  (The solid green curve shows $\Df/5$.)
  } \vspace{-5mm}
  \label{fig:XY_B}
\end{figure}

To investigate this, we revisit our first benchmark example for the domain wall motion of the XX model. We use CBE--TDVP (while exploiting $U(1)$ spin symmetry) to compute the forward-backward fidelity [Fig.~\ref{fig:XY_B}(a)]

\vspace{-6mm}

\begin{flalign}
  \label{eq:fidelity}
  & F(\tb) = |\! \braket{\Psi_-(\tb)|\Psi_+(t)}\! |^2 \,  , 
  \quad \tb  \!=\! \tmax \!-\! t  \in [0,\tmax] \, .
  \hspace{-1cm} & 
  \end{flalign}

\vspace{-1mm}
  
\noindent 
Here, $\ket{\Psi_+(t)} \!=\! e^{-\mi H t} \ket{\Psi(0)}$ is obtained 
through forward evolution for time $t$, 
and $\ket{\Psi_-(\tb)} \!=\! e^{\mi H \tb} \ket{\Psi_+ (\tmax)}$
through forward evolution until time $t\!=\! \tmax$, then back-evolution for  $\tb  \!=\!  \tmax \!-\! t$ to get back to time $t$. The deviation of the fidelity from unity,
$\delta F(\tb) \!=\! 1 \!-\! F(\tb)$,  equals zero for unitary evolution; increases with $\tb$ if time evolution is computed using truncations; and tends to 1 for $\tb \to \tmax$ if truncations are  too severe. 

Figure~\ref{fig:XY}(b) shows the back-evolution  of the domain wall described by $\ket{\Psi_-( \tb)}$  as $\tb$ increases 
from $0$ to $\tmax=40$, where both $\ket{\Psi_+(t)}$
and $\ket{\Psi_-(\tb)}$ were computed using CBE--TDVP with the truncation parameters stated in the main text, namely $\epsilont\!=\!10^{-6}$ and $\Dmax\!=\!120$.
The corresponding $\delta F(\tb)$ (Fig.~\ref{fig:XY}(d), black dashes) shows initial transient growth, but then \textit{saturates} at a remarkably
small plateau value of $6.7 \times 10^{-5}$. Moreover, the corresponding  
bond expansion per update, $\Dt(\tb)$ (Fig.~\ref{fig:XY}(e), black dots), increases only fairly slowly. For these truncation settings, the CBE--TDVP errors are thus clearly under good control and do not accumulate rapidly, so that long-time evolution can be computed accurately. 

The fidelity becomes worse ($\delta F(\tb)$ increases) if 
the singular-value threshold for bond expansion, $\epsilont$, is raised (Fig.~\ref{fig:XY}(d), dashed lines). Nevertheless,  even for $\epsilont$
as large as $10^{-2}$ we find long-time plateau behavior for $\delta F(\tb)$, implying that the errors remain controlled. This illustrates the robustness of CBE--TDVP. The plateau value can be decreased by increasing
$\Dmax$, but the reduction becomes significant only
if $\epsilont$ is sufficiently small. Even for $\Dmax \!=\! \infty$ 
(Fig.~\ref{fig:XY}(d), solid lines) the plateau reduction 
relative to $\Dmax \!=\! 120$ is modest, whereas the corresponding growth in $\Df$ (Fig.~\ref{fig:XY}(e), solid lines) becomes so rapid that this setting is not recommended in practice. 

Finally, Figs.~\ref{fig:XY}(c) and \ref{fig:XY}(d) (dash-dotted, purple line) also show \oneTDVP\ results, computed with $D=120$: the domain wall fails to recontract to a point, and the fidelity reaches zero  ($\delta F(\tb)$ reaches $1$).
This occurs even though \oneTDVP\ uses no truncations besides the tangent space projection, and hence yields unitary time evolution. 
This poor performance illustrates a key limitation of \oneTDVP\ when exploiting symmetries (as here): time evolution involves transitions to sectors having quantum numbers not yet present, but \oneTDVP\  cannot include these, due to the fixed-rank nature of its tangent space projection. CBE--TDVP by construction lifts this restriction. 

\section{Comparison of CPU time for CBE--TDVP and 2TDVP}

In this section, we compare the CPU time for CBE--TDVP and 2TDVP.
As a demonstration, we use the one-axis twisting (OAT) model discussed in \textit{Results} in the main text.
All CPU time measurements were done on a single core of an Intel Core i7-9750H processor.

\begin{figure}[t]
\includegraphics[width=1.0\linewidth,trim = 0.0in 0.0in 0.0in 0.0in,clip=true]{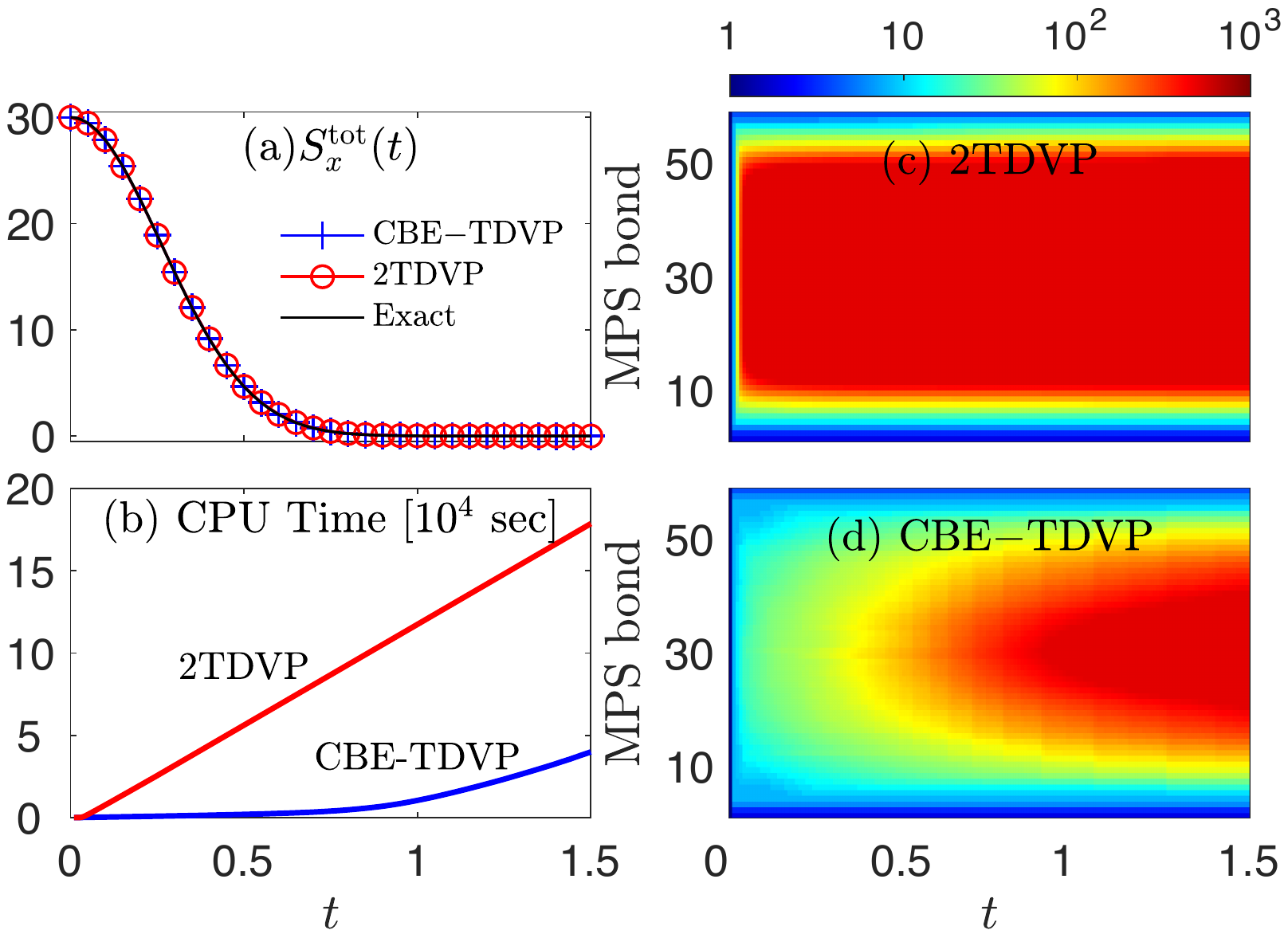}
\vspace{-6mm}
\caption{$60$-site one-axis twisting model for spin $S=1/2$: Time evolution of an initially $x$-polarized spin state, computed using $\deltat\!=\!0.01$, $\Dmax \!=\!500$, and $\mathbb{Z}_2$ spin symmetry.
  (a) Total spin $S^{\rm{tot}}_x(t)$ for CBE--TDVP (blue), 2TDVP (red) and the exact solution (black). 
  (b) CPU time for CBE-TDVP (blue) and 2TDVP (red). (c,d) Color 
  scale plot of the bond dimension as a function of time for all MPS bonds, for (c) 2TDVP and (d) CBE--TDVP. 
}
\label{fig:OAT_SI}
\vspace{-5mm}
\end{figure}

First, we compare the early-time behavior of CBE--TDVP and 2TDVP.
From $t=0$ to $1.5$, both methods yield good accuracy as shown in Fig.~\ref{fig:OAT_SI}(a).
The CPU time spent to achieve this, however, is quite different.
In Fig.~\ref{fig:OAT_SI}(b), we see that while the 2TDVP takes about two days, CBE--TDVP accomplishes the same time span overnight.

The main reason for this difference does not lie in the \onesite\ vs.\ \twosite\ scaling of CBE--TDVP vs.\ 2TDVP (discussed below),  because 
$d=2$ (for $S=1/2$) is small, and CBE involves some algorithmic overhead for determining the truncated complement $\Atrunc_\ell (\TriangleOrangeA)$. 
Instead, the difference reflects the fact that the growth in MPS bond dimension $D(t)$ with time is much slower for CBE-TDVP than 2TDVP. This implies dramatic cost savings, since both methods have time complexity proportional to $D^3$. 
Figure~\ref{fig:OAT_SI}(c,d) show the time evolution of bond dimensions for all MPS bonds for CBE--TDVP and 2TDVP respectively.
For 2TDVP [Fig.~\ref{fig:OAT_SI}(c)],  the bond dimensions grow almost exponentially and quickly saturate to their specified maximal value, here $D_{\rm{max}}=500$.
This saturation is reflected by the early onset of linear growth in the CPU time in Fig.~\ref{fig:OAT_SI}(b).
By contrast, the bond dimensions of CBE--TDVP show a much slower growth
[Fig.~\ref{fig:OAT_SI}(d)], yielding a strong reduction in CPU time compared to 2TDVP.

\begin{figure}[!h]
\includegraphics[width=1.0\linewidth,trim = 0.0in 0.0in 0.0in 0.0in,clip=true]{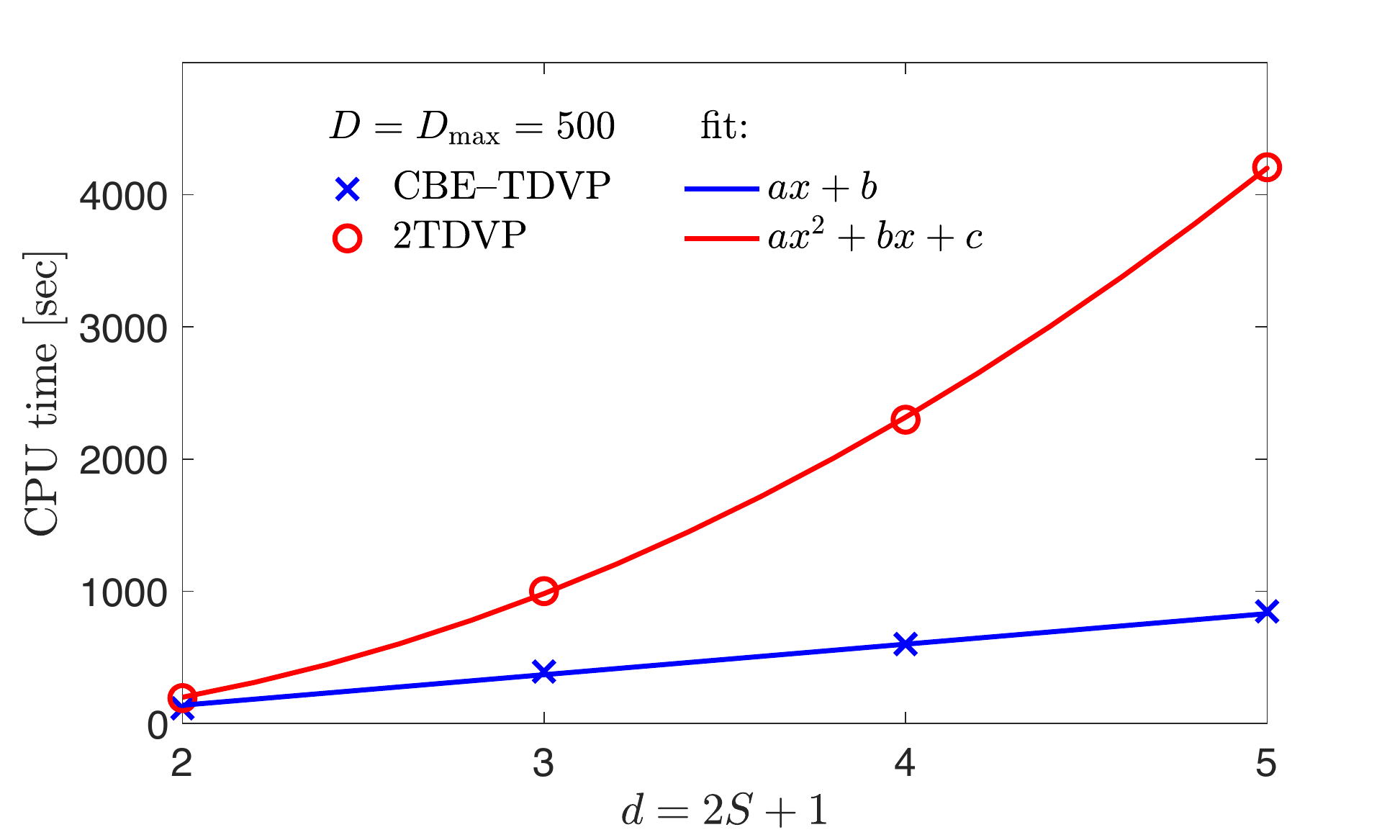}
\vspace{-6mm}
\caption{CPU time per sweep for the $20$-site one-axis twisting model, computed for several values of $S$,  at $D_{\rm{max}} = 500$. 
}
\label{fig:OAT_SI2}
\vspace{-5mm}
\end{figure}

Second, we demonstrate that when $D$ is fixed, the time complexity of CBE--TDVP vs.\ 2TDVP scales as $d$ vs.\ $d^2$, implying \onesite\ vs.\ \twosite\ scaling.
Figure~\ref{fig:OAT_SI2} shows this by displaying the 
CPU time per sweep for the OAT model for several different values of the spin $S$, with the MPS bond dimension fixed at $\Dmax= 500$.

\end{document}